# Balancing reaction-diffusion network for cell polarization pattern with stability and asymmetry

Yixuan Chen[1,2,3,†], Guoye Guan[1,4,5,†,a,b], Lei-Han Tang[1,5,6,7,*], Chao Tang[3,4,8,*]

1. *South Bay Interdisciplinary Science Center, Songshan Lake Materials Laboratory, Guangdong 523808, China*
2. *Department of Physics, Sichuan University, Sichuan 610065, China*
3. *School of Physics, Peking University, Beijing 100871, China*
4. *Center for Quantitative Biology, Peking University, Beijing 100871, China*
5. *Department of Physics, Hong Kong Baptist University, Hong Kong 999077, China*
6. *Institute of Computational and Theoretical Studies, Hong Kong Baptist University, Hong Kong 999077, China*
7. *State Key Laboratory of Environmental and Biological Analysis, Hong Kong Baptist University, Hong Kong 999077, China*
8. *Peking-Tsinghua Center for Life Sciences, Peking University, Beijing 100871, China*

† These authors contributed equally to this work.

a. Present address: *Department of Systems Biology, Harvard Medical School, Boston 02115, USA*

b. Present address: *Department of Data Science, Dana-Farber Cancer Institute, Boston 02215, USA*

* For correspondence: lhtang@hkbu.edu.hk (L.-H. Tang), tangc@pku.edu.cn (C. Tang).

**Keywords**

Reaction-diffusion network | Cell polarization | Stability | Asymmetry | *Caenorhabditis elegans* | Embryogenesis

## ABSTRACT

Cell polarization is a critical process that separates molecular species into two distinct regions in prokaryotic and eukaryotic cells, guiding biological processes such as cell division and cell differentiation. Although several underlying antagonistic reaction-diffusion networks capable of setting up cell polarization have been identified experimentally and theoretically, our understanding of how to manipulate pattern stability and asymmetry remains incomplete, especially when only a subset of network components are known. Here we present numerical results to show that the polarized pattern of an antagonistic 2-node network collapses into a homogeneous state when subjected to single-sided self-regulation, single-sided additional regulation, or unequal system parameters. However, polarity restoration can be achieved by combining two modifications with opposing effects. Additionally, spatially inhomogeneous parameters favoring respective domains stabilize their interface at designated locations. To connect our findings to cell polarity studies of the nematode *Caenorhabditis elegans* zygote, we reconstituted a 5-node network where a 4-node circuit with full mutual inhibitions between anterior and posterior is modified by a mutual activation in the anterior and an additional mutual inhibition between the anterior and the posterior. Once again, a generic set of kinetic parameters moves the interface towards either the anterior or posterior end, yet a polarized pattern can be stabilized through spatial tuning of one or more parameters coupled to intracellular or extracellular cues. A user-friendly software, *PolarSim*, is introduced to facilitate the exploration of networks with alternative node numbers, parameter values, and regulatory pathways.

## 1. Introduction

Cell polarization is a biophysical and biochemical process where a cell acquires spatial anisotropy by establishing directional gradients of molecular species across its membrane or in the cytosol [Knoblich et al., *Nat. Rev. Mol. Cell Biol.*, 2001; Goodrich et al., *Development*, 2011]. Polarity establishment is an essential step in a wide range of biological phenomena, including embryonic development, wound healing, immune activity, and so forth [Etienne-Manneville et al., *Nature*, 2002; Bilder et al. *Science*, 2000]. During cytokinesis, a polarized cell can allocate its molecular contents unequally to its daughter cells, leading to asymmetric cell size and cell fate [Macara, *Nat. Rev. Mol. Cell. Biol.*, 2004]. At the multicellular level, a group of polarized cells can undergo collective movements and constitute stereotypic architectures [Etienne-Manneville et al., *Nature*, 2002; Bilder et al. *Science*, 2000]. Thus, loss or disorder of cell polarization could severely violate normal biological processes, for example, resulting in embryonic lethality and cancerous tumor [Bilder et al., *Science*, 2000; Kim et al., *J. Cell. Sci.*, 2007]. To this day, cell polarization has been a long-term research focus in cell and developmental biology, where more and more efforts have been paid to uncover both the regulatory pathways and design principles involved [Doe et al., *Curr. Opin. Cell Biol.*, 2001; Koorman , *Nat. Cell Biol.*, 2016; Tostevin et al., *Biophys J*, 2008; Chau et al., *Cell*, 2012].

Much of our knowledge on cell polarization is derived from the zygote of the nematode *Caenorhabditis elegans* (referred to as *C. elegans*), which has served as a prominent model organism for studying cell polarization [Rose et al., *WormBook*, 2014; Lang et al., *Development*, 2017]. After fertilization, the *C. elegans* zygote is polarized where the entry side of the sperm (one pole of the ellipsoidal egg) turns into the posterior of the embryo [Goldstein et al., *Development*, 1996; Motegi et al., *Nat. Cell Biol.*, 2011]. Driven by cell polarization, the embryo proceeds tough four successive rounds of asymmetric cell divisions and produces four somatic founder cells sequentially [Sulston et al., *Dev. Biol.*, 1983; Hubatsch et al., *Nat. Phys.*, 2019].

The asymmetric division in the *C. elegans* zygote is governed by a protein family termed partitioning-defective protein (PAR) [Kemphues et al., *Cell*, 1988]. The initial PAR family only consisted of the PAR-3/PAR-6/PKC-3 complex and PAR-1/PAR-2 complex, which are stably accumulated in the anterior and posterior domains on the membrane of the zygote before its division, respectively [Cuenca et al., *Development*, 2003]. During the establishment phase triggered by the sperm entry that defines the zygote's posterior, PAR-3/PAR-6/PKC-3, initially distributed across the entire cell membrane, are anteriorly directed and localized through the contraction of the cortical actomyosin marked by non-muscle myosin II (NMY-2), which allows the PAR-2 (along with the recruitment of PAR-1) to be enriched on the posterior membrane conversely [Rose et al., *WormBook*, 2014; Lang et al., *Development*, 2017]. Subsequently, the maintenance phase follows as the cortical actomyosin contracts halfway across the zygote, with cortical NMY-2 being regulated downstream by PAR proteins and playing a minor role in maintaining the cell polarization pattern [Goehring et al., *J. Cell Biol.*, 2011; Rose et al., *WormBook*, 2014; Beatty et al., *Development*, 2013]. The cell polarization patterns, defined by the concentration distributions of PAR proteins, are stably maintained to guide the assembly of the cytokinetic ring, which is vital for the first asymmetric cell division [Goehring et al., *J. Cell Biol.*, 2011; Rose et al., *WormBook*, 2014; Jankele et al., *eLife*, 2021].

A series of experiments (including RNA interference, immunoprecipitations, fluorescence recovery after photobleaching, and time-lapse single-molecule imaging) further uncovered the mutual inhibition between PAR-3/PAR-6/PKC-3 and PAR-1/PAR-2 upon their association with the membrane and demonstrated its essential role in their robust spatial separation [Cuenca et al., *Development*, 2003; Motegi et al., *Nat. Cell Biol.*, 2011]. Soon, theoretical and numerical studies proved that mutual inhibition forms the backbone (hereafter referred to as an antagonistic 2-node network) of a polarized pattern [Tostevin et al., *Biophys J*, 2008; Chau et al., *Cell*, 2012]. In the *C. elegans* zygote, the antagonism between these two protein groups directs the uneven distribution in downstream cell fate determinants (*e.g.*, PIE-1 and P granules), which is inherited by the two daughter cells during the following asymmetric cell division, leading to cell differentiation [Schubert et al., *Mol. Cell*, 2000; Wang et al., *Adv. Exp. Med.*

*Biol.*, 2013].

In recent years, more proteins have been identified to significantly interact with the antagonistic 2-node network during the maintenance phase of *C. elegans* zygotic cell polarization, such as the CDC-42, LGL-1, and, CHIN-1 [Kumfer et al., *Mol. Biol. Cell*, 2010; Sailer et al., *Dev. Cell*, 2015; Beatty et al., *Development*, 2010]. The addition of these players increases the complexity of the cell polarity network tremendously, not only because of the explosion of the associated kinetic parameters, but also their role in related cellular processes such as cytoskeleton organization and localization at the poles that determines cell division dynamics [Ajduk et al., *Mol. Hum. Reprod.*, 2016; Lim et al., *Cell Rep.*, 2021]. Therefore, a more general understanding of the stability and asymmetry of cell polarization patterns will help to streamline experimental data analysis and interpretation. In a separate development, the theoretical knowledge acquired from nature is urgently needed for *de novo* construction of cells with designated characteristics, where the capability of cell polarization has been a design target since over a decade ago [Lin et al., *Nat. Commun.*, 2021; Watson et al., *Cell*, 2023; Tostevin et al., *Biophys J*, 2008; Chau et al., *Cell*, 2012].

In this work, we focus on the generation of stable asymmetric patterns in both the widely-used 2-node network and a more realistic *C. elegans* 5-node network. Starting from a symmetric antagonistic network, a polarized pattern can be stabilized at any interface location between the two antagonistic domains when translational symmetry is assumed. Unbalanced modification of kinetic parameters triggers movement of the interface in favor of one of the coexisting domains. There are three types of such unbalanced modification: single-sided self-regulation, single-sided additional regulation, and unequal system parameters. Nevertheless, the combination of two or more unbalanced modifications can recover pattern stability through fine-tuning of kinetic parameters. Alternatively, we show that the interface can also be stabilized at a designated location with a step-like spatial profile of one or more of the kinetic parameters, with values leading to opposing velocities when the interface is displaced. Intriguingly, such a program strategy is found to be employed in the *C. elegans* cell polarization network to maintain pattern stability along with considerable parameter robustness, while inducing pattern asymmetry by interface localization control.

## 2. Results

### *2.1. A computational pipeline for simulating cell polarization*

To investigate both the simple network and the realistic network consisting of various node numbers and regulatory pathways [Goehring et al., *Science*, 2011; Lang et al., *Development*, 2017], we propose a computational pipeline for numerical exploration of the dynamics of a given reaction-diffusion network's dynamics, specifically targeting the maintenance phase of stable cell polarization after its initial establishment [Motegi et al., *Nat. Cell Biol.*, 2011; Goehring et al., *Science*, 2011; Seirin-Lee et al., *Cells*, 2020]. Numerous biological experiments on different organisms have demonstrated that such a reaction-diffusion network typically comprises two groups of molecular species. Each group, once associated with a part of the cell membrane from the cell cytosol, inhibits the association of the other group in the same region, thereby creating two types of distinct domains [Knoblich et al., *Nat. Rev. Mol. Cell Biol.*, 2001; Doe et al., *Curr. Opin. Cell Biol.*, 2001]. At the interface between the two domains, the membrane association of proteins in either group is compromised due to the elevated level of antagonists. Nevertheless, one of the domains may expand at the expense of the other, leading to a finite interface velocity in general. For simplicity, three assumptions previously used in research are adopted to establish a reaction-diffusion model to describe the dynamics of each molecular species (denoted by [X]) during cell polarization:

1. The cellular space is reduced to a one-dimensional line of length $L = 0.5$, where $x = -0.25$ (anterior, where the concentration of molecular species accumulated on the cell membrane is denoted by $[\mathrm{A_m}]$) and $0.25$ (posterior, where the molecular species accumulated on the cell membrane is denoted by $[\mathrm{P_m}]$) are its two poles (Fig. 1a) [Gross et al., *Nat. Phys.*, 2019; Seirin-Lee, *Dev. Growth Differ.*, 2020].
2. A molecular species of type [X] can associate with the cell membrane from cell cytosol at a rate $F_{\mathrm{on}}^{\mathrm{X}}(x,t)$ and dissociate from the cell membrane into cell cytosol at a rate $F_{\mathrm{off}}^{\mathrm{X}}(x,t)$, where $t$ represents time. Both rates include leaky term and regulatory pathways affected by other molecular species [Seirin-Lee et al., *J. Theor. Biol.*, 2015; Seirin-Lee et al., *Cells*, 2020].
3. Based on previous experimental measurements, it has been reported that the diffusion rate of molecular species involved in cell polarization (*i.e.*, PAR-2 and PAR-6) is two orders of magnitude higher in the cell cytosol compared to the cell membrane [Goehring et al., *J. Cell Biol.*, 2011; Lim et al., *Cell Rep.*, 2021]. Consequently, we adopted $2.748\times10^{-5}$ and $1.472\times10^{-5}$ from previous research (nondimensionalized based on experimental measurements [Goehring et al., *J. Cell Biol.*, 2011; Goehring et al., Science, 2011; Seirin-Lee et al., *Cells*, 2020], as diffusion coefficients of anterior and posterior molecular species, while the cytoplasmic molecules are regarded as well-mixed with infinite diffusion [Kravtsova et al., *Bull. Math. Biol.*, 2014; Gross et al.,

*Nat. Phys.*, 2019; Morita et al., *J. Math. Biol.*, 2021]. We further assume that these molecular species are sufficiently abundant in the pool of cell cytosol, where the concentration is conserved as constant $[\mathrm{X_c}]$, although the analysis below can be extended to the more general case [Kravtsova et al., *Bull. Math. Biol.*, 2014; Goehring et al., *Science*, 2011].

For the concentration of each molecular species [X] on the cell membrane $[\mathrm{X_m}]$, the evolution equation is described by:

$$\frac{\partial[\mathrm{X_m}]}{\partial t} = D_{\mathrm{m}}^{\mathrm{X}} \frac{\partial^2}{\partial x^2}[\mathrm{X_m}] + F_{\mathrm{on}}^{\mathrm{X}}(x,t)[\mathrm{X_c}] - F_{\mathrm{off}}^{\mathrm{X}}(x,t)[\mathrm{X_m}], \tag{1}$$

where $D_{\mathrm{m}}^{\mathrm{X}}$ is the diffusion coefficient of molecular species [X] on the cell membrane. The first term on the right side represents the diffusion across the cell membrane; the second term represents the association with the cell membrane from the cytosol; while the third term represents the dissociation from the cell membrane into the cytosol. These regulated association and dissociation are given by the Hill equation [Seirin-Lee et al., *J. Theor. Biol.*, 2015; Seirin-Lee, *Bull. Math. Biol.*, 2021]:

$$F_{\mathrm{on}}^{\mathrm{X}}(x,t) = \gamma_{\mathrm{X}} + \sum_{\mathrm{Y}} \frac{q_{\mathrm{Y2}}^{\mathrm{X}}[\mathrm{Y_m}]^{n_{\mathrm{Yq}}^{\mathrm{X}}}}{1 + q_{\mathrm{Y1}}^{\mathrm{X}}[\mathrm{Y_m}]^{n_{\mathrm{Yq}}^{\mathrm{X}}}}, \tag{2}$$

$$F_{\mathrm{off}}^{\mathrm{X}}(x,t) = \alpha_{\mathrm{X}} + \sum_{\mathrm{Y}} \frac{k_{\mathrm{Y2}}^{\mathrm{X}}[\mathrm{Y_m}]^{n_{\mathrm{Yk}}^{\mathrm{X}}}}{1 + k_{\mathrm{Y1}}^{\mathrm{X}}[\mathrm{Y_m}]^{n_{\mathrm{Yk}}^{\mathrm{X}}}}, \tag{3}$$

where $F_{\mathrm{on}}^{\mathrm{X}}$ consists of a basal on-rate (*i.e.*, association rate) $\gamma_{\mathrm{X}}$ and recruitment by other membrane-bound molecular species ($\mathrm{Y} \neq \mathrm{X}$) and itself ($\mathrm{Y} = \mathrm{X}$, *i.e.*, self-activation), quantified by positive parameters $q_{\mathrm{Y}i}^{\mathrm{X}}$ ($i = 1, 2$); $F_{\mathrm{off}}^{\mathrm{X}}$ consists of a basal off-rate (*i.e.*, dissociation rate) $\alpha_{\mathrm{X}}$ and exclusion by other molecular species ($\mathrm{Y} \neq \mathrm{X}$) and itself ($\mathrm{Y} = \mathrm{X}$, *i.e.*, self-inhibition), quantified by positive parameters $k_{\mathrm{Y}i}^{\mathrm{X}}$ ($i = 1, 2$); the Hill coefficients $n_{\mathrm{Yq}}^{\mathrm{X}}$ and $n_{\mathrm{Yk}}^{\mathrm{X}}$ are set to 2 as used before [Seirin-Lee et al., *J. Theor. Biol.*, 2015; Seirin-Lee, *Bull. Math. Biol.*, 2021]. To simplify the model for numerical exploration with an affordable computational cost, we set the same responsive concentration $q_{\mathrm{Y1}}^{\mathrm{X}} = k_{\mathrm{Y1}}^{\mathrm{X}} = k_1$ for both activation and inhibition pathways; the inhibition intensities $k_{\mathrm{Y2}}^{\mathrm{X}}$ are set to 1 as a numerically normalized value. We also set self and mutual activation parameters $q_{\mathrm{Y2}}^{\mathrm{X}} = q_2$ to be the same for synergistic proteins in the same group. Finally, the basal association and dissociation rates are regarded the same so that these spatially-independent effects can be neutralized, in other words, $\gamma_{\mathrm{X}} = \alpha_{\mathrm{X}} = \gamma$. Consequently, there are only three independent parameters governing this system: $\gamma$, $k_1$, and $q_2$. A detailed description of the parameters is listed in Table S1. This dimensionally reduced parameter configuration is sufficient to describe the temporal evolution of polarization distribution on the membrane under the well-mixed cytoplasmic protein concentration. All the simplified parameter value assignments above will be extensively explored by giving different values to different molecular species and pathways as well as setting a heterogeneous spatial distribution.

Simulations are performed by systematically scanning the dimensionless parameter set $\Omega(\gamma, k_1, q_2)$ on a three-dimensional (3D) grid $\gamma \in [0,0.05]$ in steps $\Delta\gamma = 0.001$, $k_1 \in [0,5]$ in steps $\Delta k_1 = 0.05$, and $q_2 \in [0,0.05]$ in steps $\Delta q_2 = 0.001$. The derived nondimensionalized parameter attains values on the same order of magnitude as those observed in reality, confirming the fidelity of the proposed model in representing the real system (Supplemental Text). When it comes to a statement of switch from pattern stability to instability, the three-parameter simplification will be removed to free the parameter value assignments (*i.e.*, uniform random variables $\gamma \in [0,0.05], \alpha \in [0,0.05], k_1 \in [0,5], , k_2 \in [0,5],\ q_1 \in [0,5], q_2 \in [0,0.05]$, and $[\mathrm{X_c}] \in [0.05,5]$ added for setting up a network with cell polarization pattern stability, searched by the Monte Carlo method) concerning all nodes and regulations, auxiliarily validating the conclusion [Fishman, *Springer New York*, 1995].

Focusing on the stability and asymmetry of cell polarization during the maintenance phase, we simplify and set the initial distribution of molecular species on the cell membrane, $[\mathrm{X_m}](x, 0)$, which is polarized during the establishment phase through active actomyosin contractility and flow [Munro et al., *Dev. Cell*, 2004; Kravtsova et al., *Bull. Math. Biol.*, 2014], using the sigmoid function:

$$\begin{cases} [\mathrm{X_m}](x,0) = 1 - \dfrac{1}{1+e^{-20x}}, & \text{for anterior molecular species} \\ [\mathrm{X_m}](x,0) = \dfrac{1}{1+e^{-20x}}, & \text{for posterior molecular species} \end{cases} . \tag{4}$$

Then the set of partial differential equations (1) evolves for 500 steps with a time step of 1 (Fig. S1). The solution at $t = 500$ is saved for further analysis if:

(i) All the molecular species still have a polarized pattern, as defined by $[\mathrm{X_m}](x = -L/2, t = 500) > 0.5$ and $[\mathrm{X_m}](x = L/2, t = 500) < 0.5$ for the anterior molecular species and $[\mathrm{X_m}](x = -L/2, t = 500) < 0.5$ and $[\mathrm{X_m}](x = L/2, t = 500) > 0.5$ for the posterior molecular species.

(ii) The cell polarization pattern is stable or nearly intact over time for each molecular species, as defined by

$$\left|\frac{[\mathrm{X_m}](x_0,\ t=500)-[\mathrm{X_m}](x_0,t=499)]}{[\mathrm{X_m}](x_0,t=499)}\right| < 10^{-4}, x_0 \in [-0.25,0.25] \quad (5)$$

Implemented on *Matlab* 2022b [The MathWorks Inc., 2022b], such a computational pipeline can describe the spatiotemporal dynamics of any reaction-diffusion network design, for evaluating its capability of maintaining cell polarization (Fig. S1). Meanwhile, we construct an automatic software, *PolarSim*, for users to explore networks with arbitrarily set node numbers, parameter values, and regulatory pathways (Supplemental Text).

### *2.2. An unbalanced network structure or parameter leads to the collapse of a polarized pattern.*

Previous experimental and theoretical discoveries have uncovered the mutual inhibition between two molecular species as a fundamental design capable of generating cell polarization [Kemphues et al., *Cell*, 1988; Cuenca et al., *Development*, 2003; Tostevin et al., *Biophys. J.*, 2008; Chau et al., *Cell*, 2012]. Thus, we utilize the completely symmetric antagonistic 2-node network to investigate the behavior of its cell polarization pattern when the network structure or parameter is modified. The two nodes (*i.e.*, molecular species) placed in the anterior and posterior are denoted by [A] and [P]. As there is no activation here then $q_2$ is not considered, the computational pipeline above establishes a total of 122 viable parameter sets $(\gamma, k_1)$ (~2.37% among 5,151 sets) that can achieve stable cell polarization. We use $(\gamma = 0.05, k_1 = 0.05)$ as a representative to show how the corresponding cell polarization pattern behaves under alternative initial condition and elementary modification (Fig. 1, Fig. S2).

Regarding the initial conditions, while uniform initial distributions maintain a homogeneous pattern (Fig. S2a, b), a polarized initial distribution reliably maintains the cell polarization regardless of variations in the initial distribution, such as shifts in transition planes (Fig. S2c, d), weak polarization (Fig. S2e, f), or asymmetric curve shapes between the two molecular species (Fig. S2f, g). In contrast, an initial distribution with only Gaussian noise produces a multipolar pattern, with the two molecular species still locally excluding each other (Fig. S2h). Altogether, these findings demonstrate the necessity of additional mechanisms, such as the active actomyosin contractility and flow, for establishing initial cell polarization independently of the reaction-diffusion network during the establishment phase [Cuenca et al., *Development*, 2003; Gross et al., *Nat. Phys.*, 2019].

Regarding the elementary modification, a total of three types are exerted on the node [A]:

(i) Single-sided self-regulation (Fig. 1b): a self-activation $(q_{\mathrm{A}2}^{\mathrm{A}} = 0.012)$ or self-inhibition $(k_{\mathrm{A}2}^{\mathrm{A}} = 0.1)$ is added on the node [A].

(ii) Single-sided additional regulation (Fig. 1c and Fig. S3): a new node [L] is added in the

anterior or posterior, with mutual activation ($q_{L2}^{A}$ & $q_{A2}^{L} = 0.012$) or inhibition ($k_{L2}^{A} = k_{A2}^{L} = 0.025$) with [A]. The location of the additional node [L] decides the initial polarized pattern represented by the sigmoid function, where an anterior node starts with a high concentration in the anterior and a low concentration in the posterior, and vice versa.

(iii) Unequal system parameters (Fig. 1d): on one hand, the inhibition intensity $k_2$ from [P] to [A] is increased from 1 to 1.6; on the other hand, the cytoplasmic concentration of [A] is increased from 1 to 1.25. The other system parameters (*e.g.*, basal on-rate $\gamma$ and basal off-rate $\alpha$) will be explored independently in the next section.

In comparison with the stable cell polarization pattern generated by the basic network, the protein distribution dynamics of eight modified conditions are simulated as shown in Fig. 1 and Movie S1, where the domain of one molecular species keeps invading the domain of the other one, ending in a homogeneous distribution. Such pattern instability caused by unbalanced network modification still holds when the network is initially set up using different initial conditions and parameters (*i.e.*, $\gamma$, $\alpha$, $k_1$, $k_2$, and $[X_c]$) assigned with various values concerning all nodes and regulations (Fig. S4, Fig. S5 - 1st row, and Table S1). This intuitively reveals that the antagonistic 2-node reaction-diffusion network is prone to become unstable with its interface keeps moving when the perturbation is introduced to the network structure or parameter.

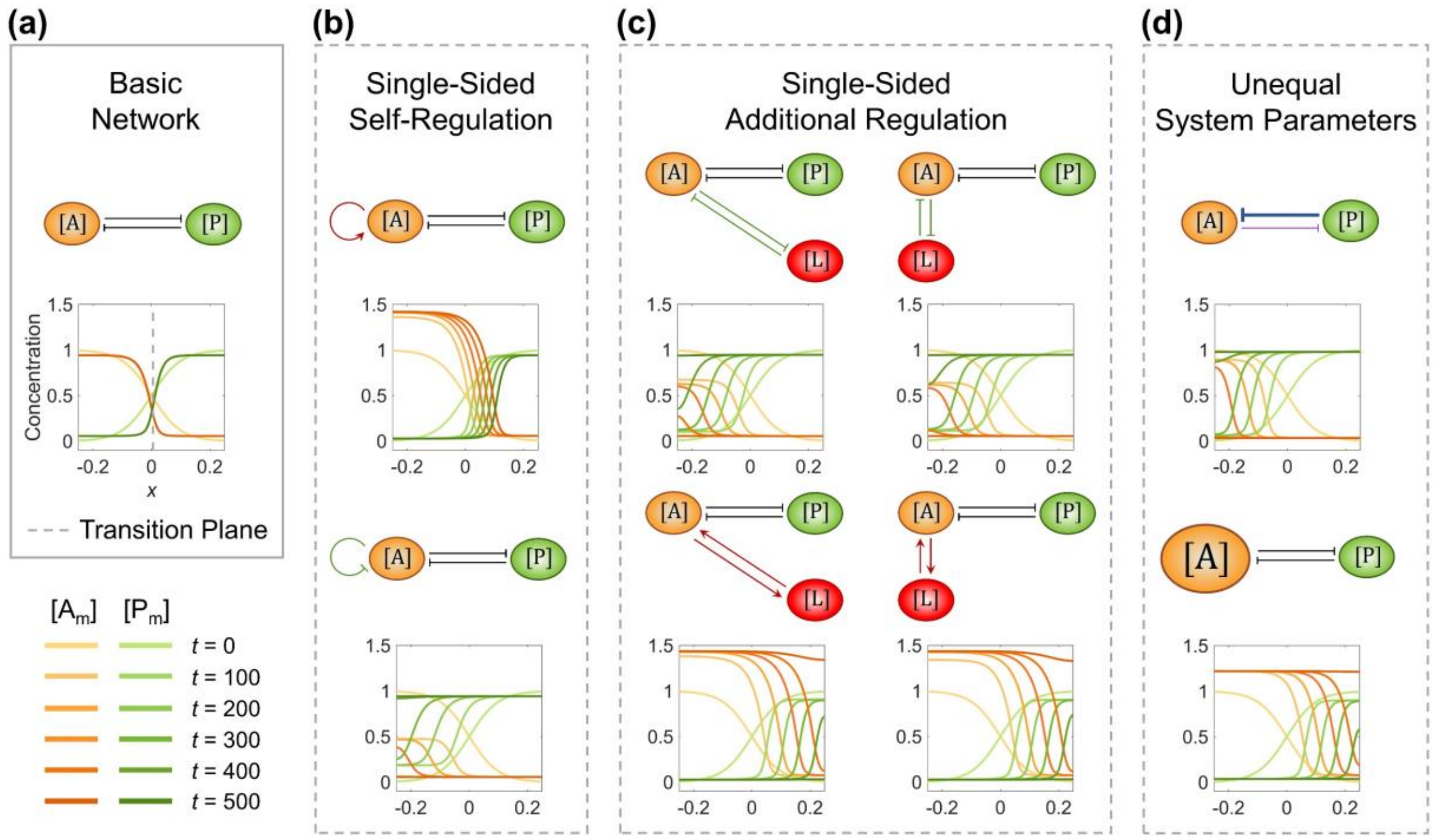

**Fig. 1.** The single modification on the antagonistic 2-node network causes the collapse of the cell polarization pattern. (a) Basic network with the transition plane marked by grey dashed line. (b) Two subtypes of single-sided self-regulation. (c) Four subtypes of single-sided additional

regulation. The corresponding spatial concentration distribution of $[L_m]$ at $t = 0, 100, 200, 300, 400,$ and $500$, and the subtle difference between left and right panels are detailed in Fig. S3. (d) Two subtypes of unequal system parameters, exemplified by unequal inhibition intensity and unequal cytoplasmic concentration. For each network, the corresponding spatial concentration distribution of $[A_m]$ and $[P_m]$ at $t = 0, 100, 200, 300, 400,$ and $500$ are shown beneath with a color scheme listed in the bottom left corner. Note that within a network, normal arrows and blunt arrows symbolize activation and inhibition respectively.

### *2.3. The combination of two modifications can recover the cell polarization pattern stability.*

Since a single modification of reaction-diffusion network structure or parameter is enough to break the cell polarization pattern, an appealing question just comes up: how can the network be designed to maintain pattern stability considering such modifications? This is crucial for cell polarization in reality where stability is essential for cell function and survival; without stability, the spatial information defined by the pattern is inaccurate for guiding downstream biological events such as cell division and cell differentiation [Hubatsch et al., *Nat. Phys.*, 2019; Wang et al., *Adv. Exp. Med. Biol.*, 2013]. The simplest idea for stability recovery is to combine two kinds of modifications with opposite trends, for example, adding self-activation and posteriorly inhibition on [A] simultaneously. For the three types of modification, we arbitrarily select one subtype within each of them from Fig. 1b-d so three combinatorial networks are constituted, all of whose interfaces turn out to be stabilized finally (Fig. 2 and Movie S2).

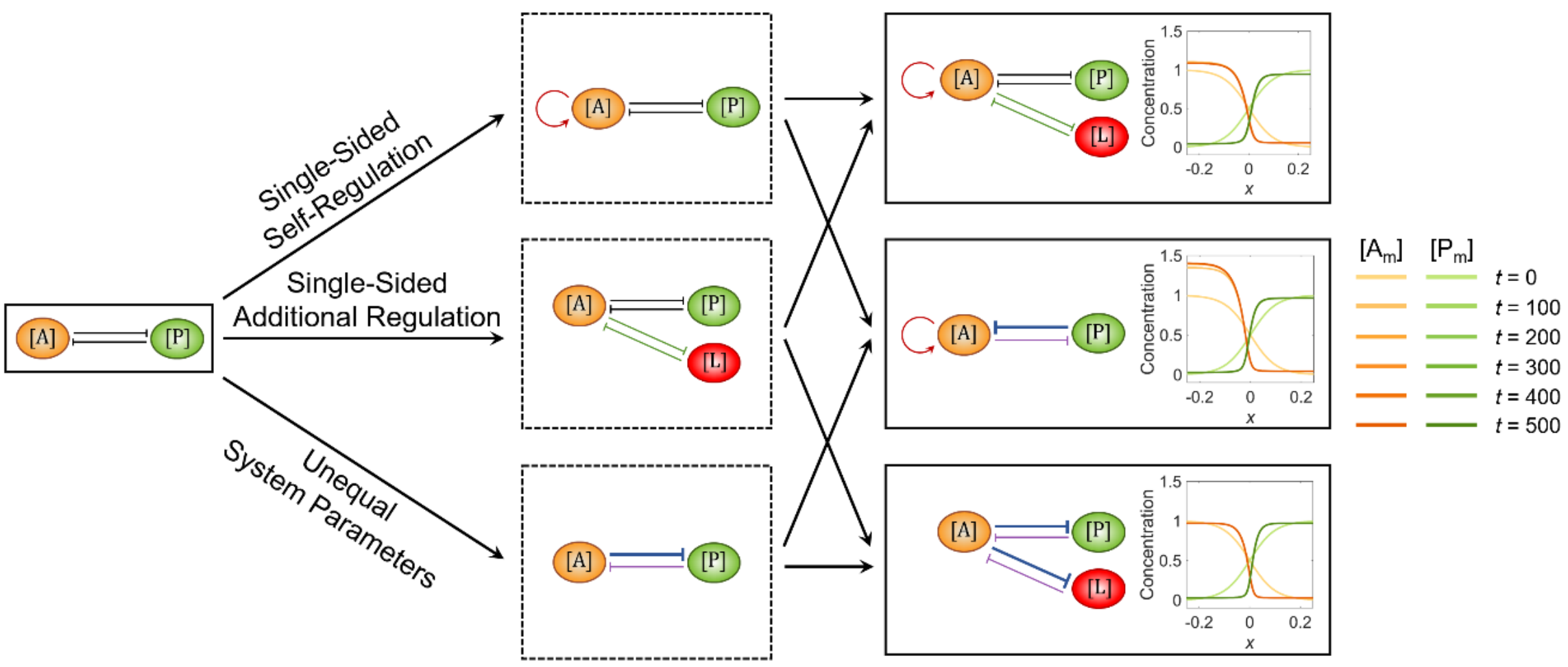

**Fig. 2.** The combination of two opposite modifications recovers the stability of the cell polarization pattern. The basic network and the ones added with a single modification are shown in the 1st and 2nd columns respectively; the three combinatorial networks composed of any two

of the three single modifications are shown in the 3rd column. For each network, the corresponding concentration distribution of $[\mathrm{A_m}]$ and $[\mathrm{P_m}]$ at $t = 0, 100, 200, 300, 400,$ and $500$ are shown beneath with a color scheme listed on the right. Here, the value assignments on the modifications in the 3rd column are as follows: $q_{\mathrm{A2}}^{\mathrm{A}} = 0.012$ and $k_{\mathrm{L2}}^{\mathrm{A}}$ & $k_{\mathrm{A2}}^{\mathrm{L}} = 0.01$ for 1st row, $q_{\mathrm{A2}}^{\mathrm{A}} = 0.012$ and $k_{\mathrm{P2}}^{\mathrm{A}} = 1.24$ for 2nd row, and $q_{\mathrm{A2}}^{\mathrm{A}} = 0.012$ and $k_{\mathrm{A2}}^{\mathrm{P}}$ & $k_{\mathrm{A2}}^{\mathrm{L}} = 2$ for 3rd row. Note that within a network, normal arrows and blunt arrows symbolize activation and inhibition respectively.

To more precisely elucidate how two opposite modifications coordinate the pattern stability together, we make use of the network with self-activation and additional inhibition on [A] (shown in the top right corner of Fig. 2), where the two corresponding intensities $q_{\mathrm{A2}}^{\mathrm{A}}$ and $k_{\mathrm{L2}}^{\mathrm{A}}$ & $k_{\mathrm{A2}}^{\mathrm{L}}$ compose a phase diagram that distinguishes the final state of the reaction-diffusion pattern. The moving velocity of the pattern is defined as follows:

$$v(t) = \frac{1}{N}\sum_{\mathrm{X}} \frac{\int_{-\frac{L}{2}}^{\frac{L}{2}} |[\mathrm{X_m}](x,t) - [\mathrm{X_m}](x,t-1)|\, \mathrm{d}x}{L}. \tag{6}$$

where $N$ is the total number of the molecular species. Here, we classify the final state by calculating the moving velocity of the pattern at the final time $t = 500$; the region with $v < 10^{-4}$ around the diagonal line is marked as the polarized state, while the regions upon and beneath it are homogeneous states dominated by [A] and [P] respectively (Fig. 3a). Therefore, the triphase diagram and the exemplary patterns generated by the parameter assignments ①② ③ shown in Fig. 3a suggest the necessity of the balance between two opposite modifications for setting up a stable cell polarization pattern.

Apart from the unequal inhibition intensity and cytoplasmic concentration mentioned in Fig. 2d, we further ask if all the kinetic parameters with biophysical significance (including inhibition intensity $k_2$, responsive concentration $k_1$, basal on-rate $\gamma$, basal off-rate $\alpha$, and cytoplasmic concentration $[\mathrm{X_c}]$) also require a balance for achieving pattern stability. For this purpose, we generate the phase diagrams between each of them and the inhibition intensity $k_{\mathrm{A2}}^{\mathrm{P}}$. Fascinatingly, monotonic correlations exist between all those system parameters, suggesting that they can be tuned to maintain pattern stability (Fig. 3b). In the grey region, the symmetric-broken parameters can be weighed against each other to realize the interface velocity close to zero.

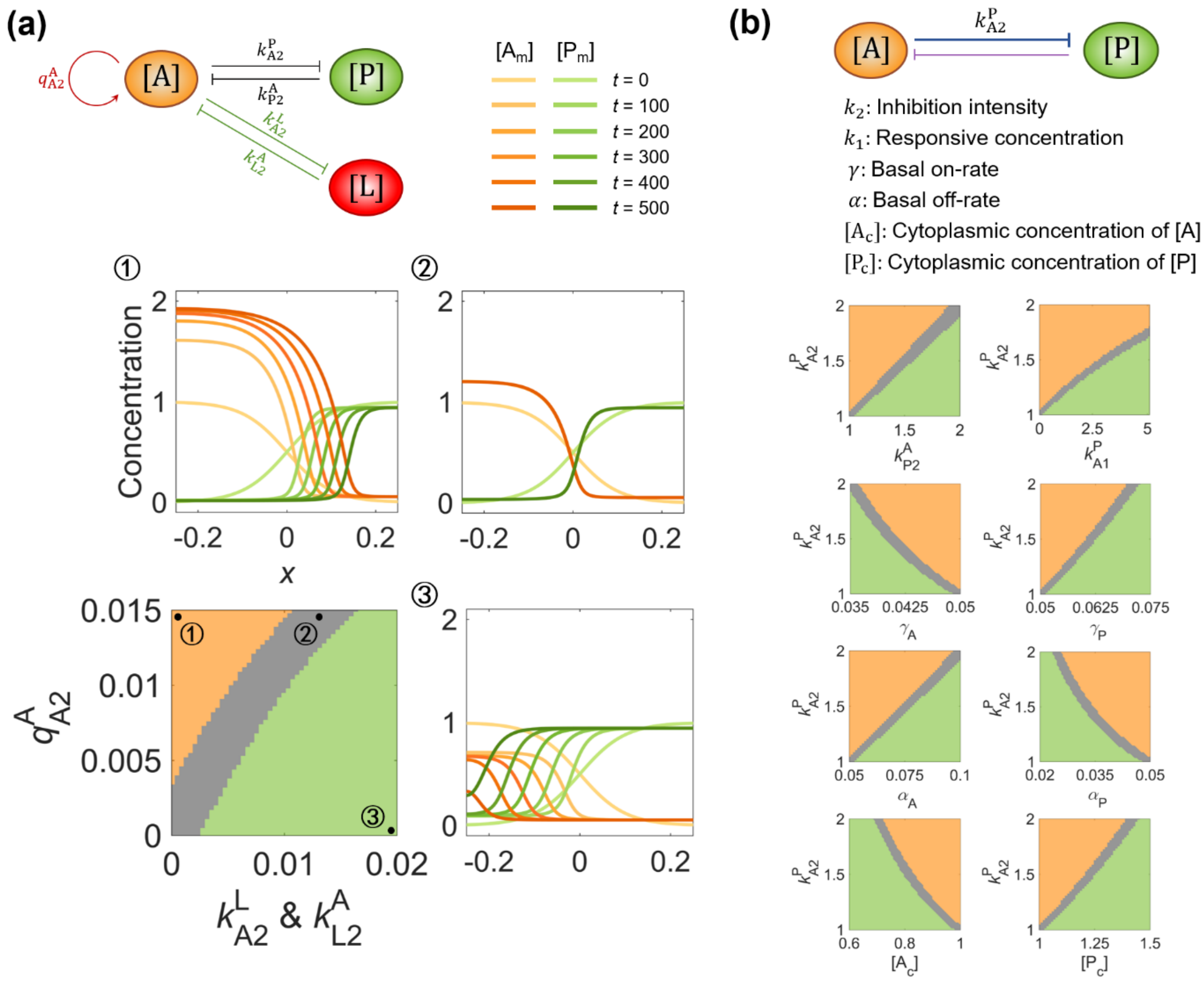

**Fig. 3.** The balance between system parameters is needed for maintaining pattern stability. (a) The phase diagram between $q_{\mathrm{A2}}^{\mathrm{A}}$ and $k_{\mathrm{L2}}^{\mathrm{A}}$ & $k_{\mathrm{A2}}^{\mathrm{L}}$ in the network modified by self-activation (quantified by $q_{\mathrm{A2}}^{\mathrm{A}}$ ) and additional inhibition (quantified by $k_{\mathrm{L2}}^{\mathrm{A}}$ & $k_{\mathrm{A2}}^{\mathrm{L}}$ ) on [A]. The representative parameter assignment for each phase are marked with ① (*i.e.*, $q_{\mathrm{A2}}^{\mathrm{A}} = 0.015$ and $k_{\mathrm{L2}}^{\mathrm{A}}$ & $k_{\mathrm{A2}}^{\mathrm{L}} = 0$ with a homogeneous state dominated by [A]), ② (*i.e.*, $q_{\mathrm{A2}}^{\mathrm{A}} = 0.015$ and $k_{\mathrm{L2}}^{\mathrm{A}}$ & $k_{\mathrm{A2}}^{\mathrm{L}} = 0.0135$ with a stable polarized state), and ③ (*i.e.*, $q_{\mathrm{A2}}^{\mathrm{A}} = 0.02$ and $k_{\mathrm{L2}}^{\mathrm{A}}$ & $k_{\mathrm{A2}}^{\mathrm{L}} = 0$ with a homogeneous state dominated by [P]). The corresponding concentration distribution of $[\mathrm{A_m}]$ and $[\mathrm{P_m}]$ at $t = 0, 100, 200, 300, 400,$ and $500$ are shown around the phase diagram with a color scheme listed on top. (b) The phase diagram between responsive concentration $k_1$, basal on-rate $\gamma$, basal off-rate $\alpha$, cytoplasmic concentration $[\mathrm{X_c}]$, and inhibition intensity $k_2$. For each phase diagram in (a)(b), the final state dominated by [A] or [P] or stable polarized is colored in orange, green, and gray, respectively. Note that within a network, normal arrows and blunt arrows symbolize activation and inhibition respectively.

### 2.4. *The velocity and position of the interface can be adjusted by setting up spatial cues.*

The required balance between parameters provides insight into controlling the pattern stability. However, in biological organisms, there exists influence from cellular actomyosin flow and extracellular/intercellular signals, leading to changeable molecular concentration and parameter values in different parts of space [Arata et al., *Development*, 2010; Beatty et al., *Development*, 2013]. How do cells maintain cell polarization pattern stability in the case of nonuniform initial conditions and parameter values, namely finding zero-velocity solutions at interfaces? Here, given a specific pattern, its position of transition plane or interface $x_{\mathrm{T}}(t)$ is defined by the mean position of all the molecular species where the absolute derivative value of the concentration curve reaches its maximum.

First, based on the symmetric 2-node network ($\gamma = 0.05, k_1 = 0.05$, Fig. 4a, the same simulation as the one in Fig. 1a), we alter its initial condition by shifting the interface position or exerting a homogeneous or Gaussian noise distribution without any positional information (Fig. S2); it is found that the initial polarized pattern is essential for constricting only one single interface with predetermined position, while the homogeneous or Gaussian noise distribution results in no or multiple interfaces (Fig. S2a, b, h). This indicates the importance of additional mechanisms (*e.g.*, the cortical flow on the cell membrane that transports proteins related to cell polarity) establishing the initial cell polarization pattern before the pattern enters the maintenance phase, when the network structure and parameters account for its stability in the long term [Cuenca et al., *Development*, 2003; Motegi et al., *Nat. Cell Biol.*, 2011].

Next, to probe into how nonuniform biophysical parameter values can control the interface in concert, we employ the following adjustment on the parameter set used in Fig. 4a:

(i) Increasing the inhibitory intensity of [A] on [P], $k_{\mathrm{A2}}^{\mathrm{P}}$, from 1 to 1.5, leads to the interface continuously shifting to the posterior, and [A] finally dominates.

(ii) Reducing the basal on-rate of [A], $\gamma_{\mathrm{A}}$, from 0.05 to 0.01, results in the interface moving to the anteriority, and [P] finally dominates.

When these two systems are combined, the stable polarity pattern remains with (i) used in the region $x < 0$ while (ii) used in $x > 0$ (Fig. 4b). By moving the step position of the parameter function to $x = -0.1$ (Fig. 4c) and $x = 0.1$ (Fig. 4d), the interface stabilizes around the step, which means the interface position is tunable (Movie S3). This indicates that by using parameter sets with values corresponding to opposite interface velocities on two sides of the interface, the interface velocity and position of the polarity pattern can be precisely regulated. The nonuniform spatial distribution of parameters imitates signaling perception, which provides a strategy for more robust control of the adjustable interface position in response to variable cues.

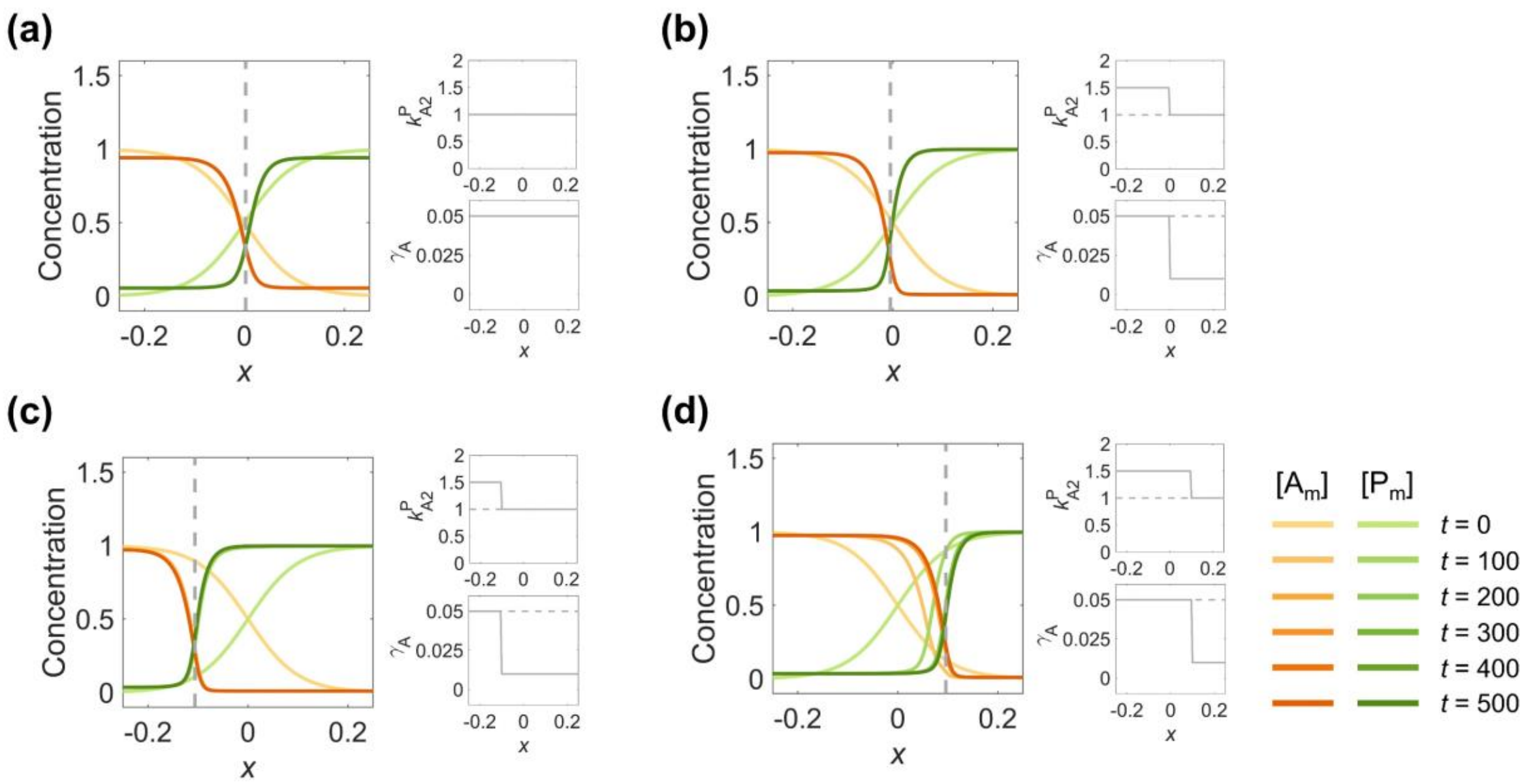

**Fig. 4.** Adopting parameter sets corresponding to opposite interface velocities on two sides of the interface, the zero-velocity interface position of the polarity pattern is shifted and regulatable. (a) Spatially uniform parameters (2nd column) of a symmetric 2-node network generate a symmetric pattern (1st column), from the same simulation in Fig. 1a. (b) Using the parameter combination with the posterior-shifting interface on the left and anterior-shifting interface on the right, a stable polarity pattern (1st column) can be remained by increasing $k_{A2}^{P}$ to 1.5 at $x < 0$ and decreasing $\gamma_A$ to 0.01 at $x > 0$ (2nd column). (c-d) The stable interface position can be optionally adjusted by setting the change position of the step-up function (2nd column). (c) As in (b), but changing the step position to $x = -0.1$, the interface stabilizes around $x = -0.1$ (1st column). (d) As in (b), but changing the step position to $x = 0.1$, the interface stabilizes around $x = 0.1$ (1st column). For each parameter set, the corresponding spatial concentration distribution of [$A_m$] and [$P_m$] at $t = 0, 100, 200, 300, 400,$ and $500$ are shown beneath with a color scheme listed on the right. Note that all the interface positions in (b-d) are marked by the vertical dashed line in gray.

### 2.5. *Reconstruction of the molecular interaction network and the design principle of parameter trade-off in C. elegans zygote.*

In the 2-node network, the cell polarization pattern stability could be broken by a single modification (Fig. 1) and recovered by two combinatorial modifications (Fig. 2, Fig. 3), or with the zero-velocity interface position regulated by spatially inhomogeneous parameters (Fig. 4). We further ask if such a fundamental rule is employed in the cell polarization network programming in a real system. To this end, we focus on the zygote of the nematode *C. elegans*, which has been a popular model for cell polarization study from both experimental and theoretical perspectives for more than 3 decades [Kemphues et al., *Cell*, 1988; Lang et al., *Development*, 2017; Goehring et al., *Science*, 2011; Lim et al., *Cell Rep.*, 2021]. By conducting an exhaustive literature search (a total of 19 references), we summarize the mutual interaction network in *C. elegans* zygote that consists of five interacting molecular species: PAR-3/PAR-6/PKC-3 complex (*abbr.*, [A]) and CDC-42 protein (*abbr.*, [C]) accumulated in the anterior, and PAR-1/PAR-2 complex (*abbr.*, [P]), LGL-1 (*abbr.*, [L]), and CHIN-1 (*abbr.*, [H]) accumulated in the posterior (Fig. 5a). The detailed description of the biochemical mechanism of each regulatory pathway as well as the corresponding supporting references are listed in Table S2. Based on the network obtained experimentally, the computational pipeline described in Section 2.1 establishes a total of 602 viable parameter sets $(\gamma, k_1, q_2)$ (~0.229% among 262,701) that can achieve stable cell polarization (Fig. 5d, e).

To verify whether the computational pipeline is reliable enough to simulate the dynamics of protein distribution of the *C. elegans* network *in vivo*, we further reproduce a series of perturbation experiments reported previously. It's worth noting that the reconstructed *C. elegans* 5-node network appears to be an integration of an antagonistic 2-node network doubled into an antagonistic 4-node network (Fig. 5b) with full connections, and the [A]~[C] mutual activation (Fig. 5c) and [A]~[L] mutual inhibition (Fig. 5d). As the antagonistic 2-node network has been revealed to be a fundamental structure capable of stable cell polarization [Tostevin et al., *Biophys J*, 2008; Goehring et al., *Science*, 2011; Chau et al., *Cell*, 2012], hereafter we focus on how the added interactions involved with [C] (the protein accumulated in the anterior and with an additional mutual activation with [A] in the anterior) and [L] (the protein accumulated in the posterior and with an additional mutual inhibition with [A] in the anterior) impact the pattern stability concordantly (Fig. S6).

Taking [L] as the first example for its current absence of theoretical model analysis [Seirin-Lee et al., *Cells*, 2020, Lim et al., *Cell Rep.*, 2021], a series of experimental groups in different conditions were conducted before by knocking down or overexpressing [L], followed by a measurement of the lethality in embryo individuals (defined by death rate and proven to be a consequence of polarity loss in the zygote, which is essential for asymmetric cell division and cell differentiation) [Beatty et al., *Development*, 2010; Beatty et al., *Development*, 2013; Rodriguez et al., *Dev. Cell*, 2017; Jankele et al., *eLife*, 2021]. Here, we utilize the pattern error in a mutant (*abbr.*, MT) embryo compared to the wild-type (*abbr.*, WT) one, $\mathrm{Error} = \frac{1}{M}\sum_{i=1}^{M}\frac{\int_{-\frac{L}{2}}^{\frac{L}{2}}|[\mathrm{X_m}]_{\mathrm{WT}}(x,t=500)-[\mathrm{X_m}]_{\mathrm{MT}}(x,t=500)|\,\mathrm{d}x}{L}$, to represent level of polarity loss in simulation and lethality in experiment (measured from previous literature), where $M = 602$ is the number of viable parameter sets (~0.229% among 262,701). The first group of experiments is that the double depletion on [P] and [L] leads to much more lethality than single depletion on either [P] or [L], which is recapitulated by Fig. S7a, b (cell polarization pattern characteristics in simulation: larger pattern error in double depletion on [P] and [L], compared to single depletion on either [P] or [L]) [Hoege et al., *Curr. Biol.*, 2010]. The second group of experiments is that the overexpression of [L] lowers the lethality induced by single depletion on [P] and such effect is weakened when [H] is depleted as well, which is recapitulated by Fig. S7c, d (cell polarization pattern characteristics in simulation: smaller pattern error in single depletion on [P] supplemented by overexpression of [L], compared to single depletion on [P] only; less smaller pattern error in double depletion on [P] and [L] supplemented by overexpression of [L], compared to single depletion on [P] only) [Beatty et al., *Development*, 2010; Beatty et al., *Development*, 2013].

The second example of theoretical model analysis involves [C]. A series of experimental groups in different conditions were conducted before by knocking down or overexpressing [C] or blocking the mutual activation between [A] and [C], followed by an observation of the pattern polarity in embryo individuals. The first group of experiments is that the single depletion on [C] results in polarity defects and the mislocalization of PAR-6 and PAR-2, which is recapitulated by Fig. S8a, b (cell polarization pattern characteristics in simulation: less polarized distributions of [A] and [P], compared to WT) [Gotta et al., *Curr. Biol.*, 2001; Aceto et al., *Dev. Biol.*, 2006]. The second group of experiments is that the overexpression of [C] increases the size of the anterior domain, which is recapitulated by Fig. S8a, c (cell polarization pattern characteristics in simulation: posteriorly shifted interfaces of [A] and [P], compared to WT) [Aceto et al., *Dev. Biol.*, 2006]. The third group of experiments is that blocking the interaction between PAR-6 and CDC-42 results in polarity defects very similar to those associated with single depletion on CDC-42, which is recapitulated by Fig. S8a, d (cell polarization pattern characteristics in simulation: less polarized distribution of [A] in both conditions, compared to WT) [Aceto et al., *Dev. Biol.*, 2006].

To sum up, apart from the cell polarization in a wild-type embryo, our modeling framework is further validated by reproducing two groups of perturbation experiments that haven't been theoretically explained before, allowing further computational investigation of the network dynamics.

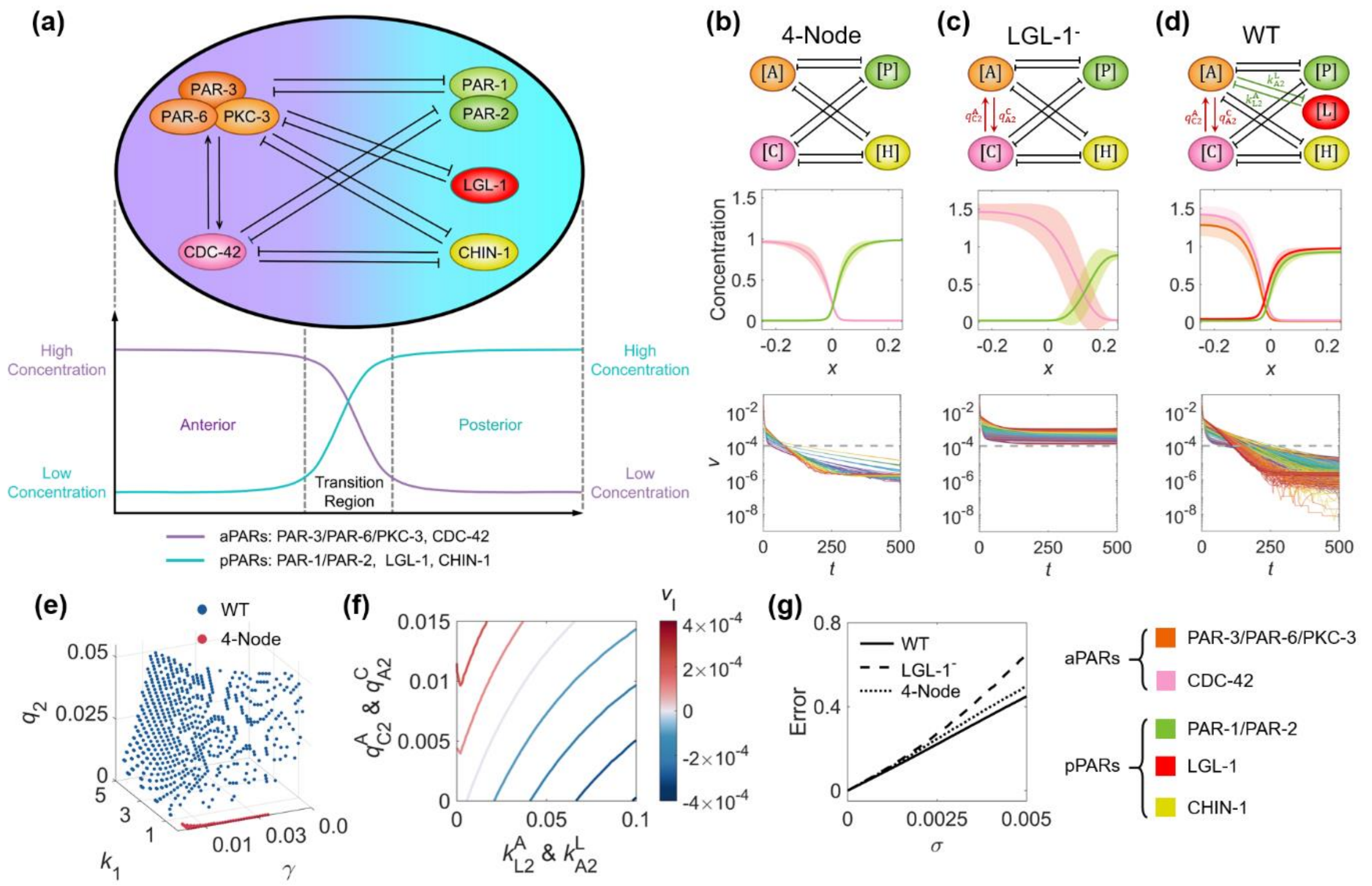

**Fig. 5.** The molecular interaction network in *C. elegans* zygote and its natural advantages in terms of pattern stability, viable parameter sets, balanced network configuration, and parameter robustness. (a) The schematic diagram of the network composed of five molecular species, each of which has a polarized spatial concentration distribution on the cell membrane shown beneath. The left nodes PAR-3/PAR-6/PKC-3 (*i.e.* [A]) and CDC-42 (*i.e.* [C]) exhibit high concentration in the anterior pole and low concentration in the posterior pole, schemed by the purple line

beneath; the right nodes PAR-1/PAR-2 (*i.e.* [P]), LGL-1 (*i.e.* [L]), and CHIN-1 (*i.e.* [H]) exhibit low concentration in the anterior pole and high concentration in the posterior pole, schemed by the cyan line beneath. Note that within the network, normal arrows and blunt arrows symbolize activation and inhibition respectively. (b-d) The structure of 4-Node, LGL-1$^{-}$, and WT networks (1$^{st}$ row). The final spatial concentration distribution ($t = 500$) averaged over all established viable parameter sets for each molecular species (*i.e.*, 62 among 262,701 sets, ~0.024%, for the 4-Node network; 62 among 5,151 sets, ~1.2%, for the LGL-1$^{-}$ network; 602 among 262,701 sets, ~0.229%, for the WT network), shown by a solid line (2$^{nd}$ row). For each position, MEAN ± STD (*i.e.*, standard deviation) calculated with all viable parameter sets is shown by shadow. The moving velocity of the pattern ($v$) over evolution time ($t$) (3$^{rd}$ row). For each subfigure in 3$^{rd}$ row, each line with a unique color represents the simulation of a unique viable parameter set, and $v = 10^{-4}$ is marked by a dashed line. For (a-d), the legend for the relationship between molecular species and corresponding color is placed in the bottom right corner. (e) The viable parameter sets of WT (blue points; 602 viable solutions among 262,701 sets, ~0.229%) and 4-Node networks (red points; 62 viable solutions among 5,151 sets, ~1.20%). (f) The interplay between [A]~[C] mutual activation and [A]~[L] mutual inhibition on the interface velocity. The contour map of the interface velocity $v_{\mathrm{I}}$ with different parameter combinations of $q_{\mathrm{C2}}^{\mathrm{A}}$ & $q_{\mathrm{A2}}^{\mathrm{C}}$ and $k_{\mathrm{L2}}^{\mathrm{A}}$ & $k_{\mathrm{A2}}^{\mathrm{L}}$ represents its moving trend. The darker the red is, the faster the interface is moving toward the posterior pole; the darker the blue is, the faster the interface is moving toward the anterior pole; the closer the color is to white, the more stable the interface is. (g) The averaged pattern error in a perturbed condition compared to the original 4-Node, LGL-1$^{-}$, and WT networks. The Gaussian noise is simultaneously exerted on all the parameters $\gamma$, $\alpha$, $k_1$, $k_2$, $q_1$ and $q_2$, where standard deviation $\sigma$ denotes the noise amplification. For a specific noise level, the pattern error is averaged over 1000 independent simulations.

With the well-validated model of the *C. elegans* cell polarization network, here we study if it follows the balance design revealed by the exhaustive study on the antagonistic 2-node network. Interestingly, the central structure of the *C. elegans* network is shown to be a completely symmetric network composed of 2 nodes in the anterior (*i.e.*, [A] and [C]) and posterior (*i.e.*, [P] and [H]) respectively; this symmetric 4-node network is modified by a mutual activation in the anterior (*i.e.*, [A]~[C]) firstly and an additional mutual inhibition between the anterior and the posterior (*i.e.*, [A]~[L]) subsequently, turning into an asymmetric 5-node network (Fig. 5b-d). The abovementioned symmetric structure, the mutual activation, and the additional inhibition are in analogy with the basic 2-node network modified by self-activation and additional inhibition as shown in the top right corner of Fig. 2 and Fig. 3a.

Here, we seek to compare the concentration distribution and moving velocity of the patterns generated by the three successive network structures termed “4-Node” (the symmetric structure) (Fig. 5b), “LGL-1⁻” (the symmetric structure with [A]~[C] added, *i.e.*, the mutant network with [A]~[L] depleted from the wild-type network) (Fig. 5c), and “WT” (the symmetric structure with both [A]~[C] and [A]~[L] added, *i.e.*, the wild-type network) (Fig. 5d). As expected, both simulations on the 4-Node and WT networks successfully pass the computation pipeline described in Section 2.1 with 62 (~1.20% among 5,151) and 602 (~0.229% among 262,701) viable parameter sets respectively (Fig. 5e), all of which stabilize into a cell polarization pattern with the concentration distribution highly intact and the moving velocity of each parameter set continuously declining below $10^{-4}$ (Fig. 5b, d and Movie S4). Nonetheless, the intermediate one (generated by depleting [A]~[L] from the WT network) with [A]~[C] but without [A]~[L] fails (Fig. 5c and Movie S4), exhibiting a dispersed concentration distribution and stubbornly high moving velocity for its pattern. Such pattern instability caused by unbalanced network modification still holds when the network is initially set up using different initial conditions and parameters (*i.e.*, γ, $\alpha$, $k_1$, $k_2$, $q_1$, $q_2$ and $[X_c]$) assigned with various values concerning all nodes and regulations (Fig. S5 - 2nd row, Fig. S9, Table S1), just like the simple 2-node network (Fig. S4, Fig. S5 – 1st row, Table S1). Therefore, WT shows the largest variable parameter sets in the limited 3D grid and the strongest stability among the three network structures.

Remarkably, the balance between these two modifications is required to achieve stable cell polarization (Fig. 5f). To evaluate how the interface moves, we quantify the interface velocity by the rate of the transition plane $x_{\mathrm{T}}$ with respect to time. To avoid the influence of the initial state, the interface velocity is then calculated as the average velocity from $t = 300$ to $t = 500$.

$$v_{\mathrm{I}} = \frac{x_{\mathrm{T}}(t = 500) - x_{\mathrm{T}}(t = 300)}{200} \tag{7}$$

Scanning the intensities of [A]~[C] mutual activation (*i.e.*, $q^{\mathrm{C}}_{\mathrm{A2}}$ & $k^{\mathrm{A}}_{\mathrm{C2}}$) and [A]~[L] mutual inhibition (*i.e.*, $k^{\mathrm{L}}_{\mathrm{A2}}$ & $k^{\mathrm{A}}_{\mathrm{L2}}$), a contour map is shown in consideration of the interface velocity averaged over all viable parameter sets (*i.e.*,$v_{\mathrm{I}}$): the system can be stably polarized when they are in balance with its interface velocity close to zero, but the overshoot of either intensity leads to a homogeneous state in the end (Fig. 5f). The redder the contour color is, the faster the interface moves posteriorly, with an increasing aPARs domain; the bluer the contour color is, the faster the interface moves anteriorly, with pPARs invading.

Next, we wonder if such combinatorial modifications may be an optimal choice selected during evolution. To this end, we regard the [A]~[C] mutual activation as a primary modification that induces pattern asymmetry as proposed before [Seirin-Lee et al., *Cells*, 2020; Lim et al., *Cell Rep.*, 2021], then the existence, as well as the form of the feed loops between [L] and a preexisting molecular species, is a supporting modification that consolidates stable polarization. Considering the symmetric structure of the LGL-1- network, there is an identical role between [A] and [C] and between [P] and [H], so [L] can be effectively connected to [A] or [P]; meanwhile, there are three types of directional regulation between them: activation, inhibition, and none. Thus, in theory, there are 32 possible network structures with an additional regulation, in which [L] must interact with one of the existing nodes (Fig. S10). Strikingly, only the WT network with mutual inhibition between [A] and [L] passes the computational pipeline with viable parameter sets. Thus, without any parametric asymmetry concessions, the configuration of the *C. elegans* network in nature is well optimized among all other alternatives for maintaining cell polarization pattern stability.

As pattern stability means how fast the pattern moves over time, does the lack of pattern stability result in a more dispersed concentration distribution when the system parameters fluctuate in time? To test this hypothesis, for each viable parameter set, we exert Gaussian noise on all the original values of system parameters $\gamma$, $\alpha$, $k_1$, $k_2$, $q_1$, and, $q_2$, and initiate 1000 independent simulations, where the noise amplification is represented by the standard deviation $\sigma$ of the Gaussian noise [Guan et al., *Phys. Rev. E*, 2021]. To compare the variance of perturbed condition (*abbr.*, PT) to the original pattern (*abbr.*, OP, including, 4-Node, LGL-1-, and WT), the pattern error, averaged over all molecular species, all the viable parameter sets, and all independent simulations, is defined as follows:

$$\text{Error} = \frac{1}{MQN}\sum_{i=1}^{M}\sum_{j=1}^{Q}\sum_{\text{X}}\frac{\int_{-\frac{L}{2}}^{\frac{L}{2}}\left|[\text{X}_{\text{m}}]_{\text{OP}}(x,t=500)-[\text{X}_{\text{m}}]_{\text{PT}}(x,t=500)\right|\text{d}x}{L} \quad (8)$$

where $M, Q, N$ represents the number of the viable parameter sets (*i.e.* 602 (~0.229% among 262,701) for WT, 62 (~1.20% among 5,151) for 4-Node and 62 (~0.024% among 262,701) for LGL-1-), independent simulations (*i.e.* 1000) and molecular species (*i.e.* five for WT, four for 4-Node and LGL-1-). It turns out that the pattern error is always the smallest in the WT network and biggest in the LGL-1- network, no matter how strong the noise is, indicating parameter robustness as the companion advantage of pattern stability (Fig. 5g).

### 2.6. *A protocol to identify responsive parameters for interface positioning*

For a polarization pattern to be stable, fine-tuning of the kinetic parameters is required to reach the vanishing velocity of the pattern interface (numerically identified by Eq. (5)). Using the 5-node *C. elegans* network as an example, we outline a method to delineate iso-velocity surfaces $v_{\mathrm{I}}(\Omega) = \text{constant}$ in the high-dimensional space of the parameter set $\Omega$. We show that the information gained can be used to quantify the role of individual molecular components in controlling the polarized pattern. Additionally, this knowledge enables us to design experiments to produce desired patterns.

The method of iso-velocity surface delineation is outlined as the following. In Sec. 2.5, 602 (~0.229% among 262,701) groups of parameters capable of reaching stable polarity patterns are found by scanning through parameter space. When plotted in the three-dimensional parameter space shown in Fig. 5e, they span an iso-velocity surface (blue points in Fig. 5e) at vanishing velocity, in other words, their pattern interfaces are stabilized with little or no moving. To illustrate the local orientation of the surface, we selected 9,261 points (enumerated on a 3D grid $\gamma \in [0.034{,}0.044]$ in steps $\Delta\gamma = 0.0005$, $k_1 \in [1.05{,}2.05]$ in steps $\Delta k_1 = 0.05$, and $q_2 \in [0.045{,}0.055]$ in steps $\Delta q_2 = 0.0005$) in the neighborhood of a benchmark point $\Omega^*(\gamma^* = 0.039, k_1^* = 1.55, q_2^* = 0.05)$, as shown by the brown box in Fig. 6a, top. The least-square fit of computed interface velocity to a linear function yields:

$$f(\gamma, k_1, q_2) = -0.0308 \times (\gamma - \gamma^*) - 6.08 \times 10^{-4} \times (k_1 - k_1^*) + 0.02 \times (q_2 - q_2^*), \quad (9)$$

with the coefficient of determination:

$$R^2 = 1 - \frac{\sum_{i=1}^{9261} [v_{\mathrm{I}}(\Omega_i) - f(\Omega_i)]^2}{\sum_{i=1}^{9261} \left[v_{\mathrm{I}}(\Omega_i) - \frac{1}{9261} \sum_{j=1}^{9261} v_{\mathrm{I}}(\Omega_i)\right]^2} = 0.9918. \quad (10)$$

The coefficients in Eq. (9) give the ascending gradient of the interface velocity (Fig. 6a, bottom).

Within the assumed relationships adopted for parameter reduction (see Sec. 2.1), Eq. (9) shows a strong dependence of the interface velocity on basal on/off-rates $\gamma$ and activation intensity $q_2$, but weak sensitivity to responsive concentration $k_1$ in the inspected parameter region. Interestingly, increase in the basal on/off-rates $\gamma$ and the self-activation rate $q_2$ have opposite effects on the interface velocity, which can be attributed to the asymmetric 5-node network topology. In a more realistic scenario, one may consider on/off-rates for different molecular species and their pairwise interactions separately and explore the relationships in a much larger parameter space. A formula similar to Eq. (9), using a linear expansion around the parameter sets balanced for a zero-velocity interface, enables quantitative prediction of the interface velocity against individual or simultaneous changes of parameters, even when knowledge of the polarity circuit is incomplete. In particular, the parameter-dependent interface velocity picture potentially enables biologists to manipulate and even synthesize the cell polarization pattern by rational modulation of parameters, controlling the interface location for specific physiological functions, such as designating a desired cell volume partition ratio [Hubatsch et al., *Nat. Phys.*, 2019; Guan et al., *Phys. Rev. E*, 2021].

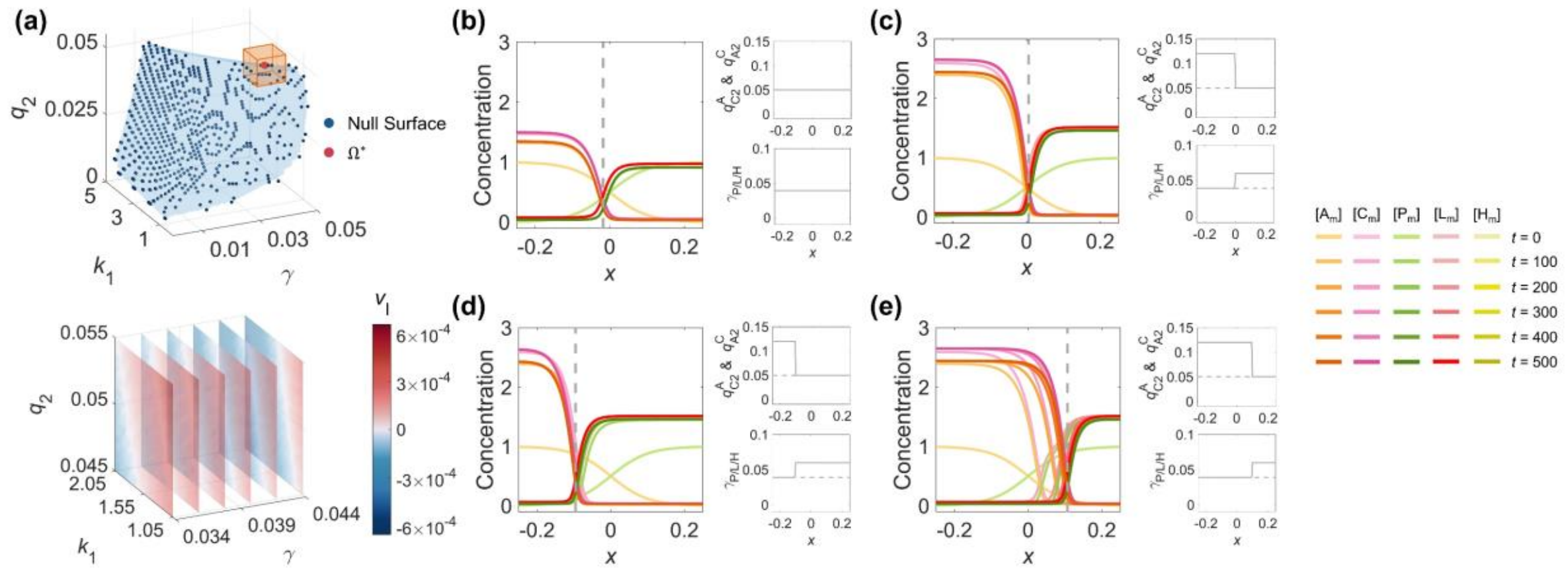

**Fig. 6.** The control of the interface velocity and position by adjusting parameters in a multi-dimensional system. (a) The parameter space (1st row) and the linear relationship (2nd row) between interface velocity and parameters. The discrete viable parameter space of WT network is fitted by a blue curved surface to represent its null surface with little or no pattern interface moving (1st row). The benchmark point $\Omega^*(\gamma^* = 0.039, k_1^* = 1.55, q_2^* = 0.05\,)$ and its neighborhood are marked by an orange cube. Centering on the benchmark point $\Omega^*$, the relationship between the velocity interface and parameters in the region marked by the orange cube is shown by slice planes orthogonal to the $\gamma$-axis at the values 0.034, 0.036, 0.038, 0.04, 0.042, and 0.044 (2nd row). The darker the red is, the faster the interface is moving toward the posterior pole; the darker the blue is, the faster the interface is moving toward the anterior pole; the closer the color is to white, the more stable the interface is. (b-d) As in Fig. 4, the control of the interface position by spatially inhomogeneous parameters corresponding to opposite interface velocities on two sides of the interface can be applied to the *C. elegans* 5-node network. (b) Using $\Omega^*$ as a representative, spatially uniform parameters (2nd column) generate a stable polarity pattern (1st column). (c) On top of (b), a stable polarity pattern (1st column) with its interface around $x = 0$ can be obtained by increasing $q_{\mathrm{C2}}^{\mathrm{A}}$ & $q_{\mathrm{A2}}^{\mathrm{C}}$ to 0.12 at $x < 0$ and increasing $\gamma_{\mathrm{P}}, \gamma_{\mathrm{L}},$ and $\gamma_{\mathrm{H}}$ to 0.06 at $x > 0$ (2nd column). (d) As in (c) but changing the step position to $x = -0.1$ (2nd column), the interface stabilizes around $x = -0.1$ (1st column). (e)

As in (c), but changing the step position to $x = 0.1$ (2nd column), the interface stabilizes around $x = 0.1$ (1st column). For each parameter set, the corresponding spatial concentration distribution of [$A_m$], [$C_m$], [$P_m$], [$L_m$], and [$H_m$] at $t = 0, 100, 200, 300, 400,$ and $500$ are shown beneath with a color scheme listed on the right. Note that all the interface positions in (b-e) are marked by the vertical dashed line in gray.

Experimental studies have revealed that the interface localization in the *C. elegans* zygotic cell polarization pattern provides an accurate spatial cue to regulate the dynamics of their downstream molecules, like the protein LIN-5 and microtubules that control cell division and volume partition [Schneider et al., *Annu. Rev. Genet.*, 2003; Ajduk et al., *Mol. Hum. Reprod.*, 2016], while the acquired cell volume has been reported to have a chain effect in its cell cycle, cell position, and other behaviors [Fickentscher et al., *New J. Phys.*, 2018; Guan et al., *Phys. Rev. E*, 2021; Fickentscher et al., *Phys. Rev. Lett.*, 2016; Tian et al., *Phys. Biol.*, 2020]. As in the 2-node network, the precise control of the stable position of the interface can be realized by taking the parameters corresponding to the opposite direction of the interface velocity on both sides of the interface. Based on the parameter set $\Omega^*$ which gives out the most stable pattern (Fig. 6b and Movie S5), two modifications are adopted to generate patterns with opposite interface velocity:

(i) Increasing the activation intensity between [A] and [C] ($q_{\mathrm{C2}}^{\mathrm{A}}$ & $q_{\mathrm{A2}}^{\mathrm{C}}$) to 0.12 leads to the interface traveling backward, and [A] and [C] finally dominate the whole domain of the membrane.

(ii) Increasing the basal on-rate of posterior proteins ($\gamma_{\mathrm{P}}, \gamma_{\mathrm{L}}, \gamma_{\mathrm{H}}$) to 0.06 results in the interface traveling forward, and [P], [L] and [H] finally dominate the whole domain of the membrane.

Combining the two sets of parameters with its step switch located at $x = 0$ ((i) on the left and (ii) on the right), the stabilized polarity interface settles around $x = 0$ (Fig. 6c and Movie S5). The interface position turns tunable as the step switch moves to $x = -0.1$ (Fig. 6d and Movie S5) and $x = 0.1$ (Fig. 6e and Movie S5). One pre-known example in reality that matches this scheme is the asymmetric division of the P2 and P3 cells (the granddaughter and great-granddaughter cells of the *C. elegans* zygote P0), where the extracellular protein MES-1 and intracellular protein SRC-1 transduct signals from the EMS and E cells (the sister cell of P2 and its posterior daughter cell), inducing polarity reversal by affecting the on-rate of PAR-2 [Arata et al., *Development*, 2010; Seirin-Lee, *J. Theor. Biol.*, 2016]. Such a scheme depicts the introduction of extracellular or intracellular cues that break the spatial uniformity of parameters and tune the stable localization of interface, serving as a theoretical basis for controlling oriented cell division with designated volume partition and fate differentiation [Arata et al., *Development*, 2010; Hubatsch et al., *Nat. Phys.*, 2019; Schubert et al., *Mol. Cell*, 2000].

## 3. Discussion and conclusion

Cell polarization is a fundamental issue in both prokaryotes and eukaryotes, playing crucial roles in diverse biological phenomena ranging from chemotaxis to embryogenesis [Nance, *J. Cell Biol.*, 2014; Kondo et al., *J. Leukoc. Biol.*, 2019]. The reaction-diffusion network responsible for cell polarization has been a long-term research focus for both experimentalists and theorists, who have identified many interactive functional molecules, discovered underlying design principles for networks, and even synthesized new systems *de novo* [Koorman et al., *Nat. Cell Biol.*, 2016; Lang et al., *Development*, 2017; Tostevin et al., *Biophys J*, 2008; Chau et al., *Cell*, 2012; Lin et al., *Nat. Commun.*, 2021; Watson et al., *Cell*, 2023]. In this paper, we focus on a basic problem – how to control the cell polarization pattern stability over time and regulate the stabilized interface localization, from the perspective of a reaction-diffusion network. First, we established a computational pipeline to search the viable parameter sets in an $N$-node (molecular species) network that can achieve a stable cell polarization pattern. The simple antagonistic network with only two nodes was revealed to be unstable (*i.e.*, transit from a polarized distribution to a homogeneous distribution spontaneously) when any of the 3 unbalanced modifications (*i.e.*, single-sided self-regulation, single-sided additional regulation, and unequal system parameters) were introduced. To recover stable polarization, two strategies are proposed: (i) the combination of two unbalanced modifications with opposite effects; (ii) the spatially inhomogeneous parameter values, either of which can lead to opposite interface velocity. Additionally, the stable interface localization can be discretionarily regulated by the step-like parameter profile and contributes to the asymmetric geometry of the cell polarization pattern, potentially providing a spatial cue for significant physiological functions like unequal cell volume partition and consequent punctuated cell movement [Fickentscher et al. *New J. Phys.*, 2018; Guan et al., *Phys. Rev. E*, 2021; Fickentscher et al., *Phys. Rev. Lett.*, 2016; Tian et al., *Phys. Biol.*, 2020]. Analogous conclusions are further identified in the *C. elegans* network summarized experimentally which obtains larger parameter space and is higher robustness against parameter perturbations, supporting them as strategies indeed applied in real biological scenarios. Importantly, the linear relationship between the interface velocity and biophysical parameters serves as a useful tool to characterize and predict how the cell polarization pattern, especially where the interface is located or moved toward, responds to parameter changes; this not only helps understand the regulatory molecular species and pathways, as well as the biophysical parameters learned from experiments, play their role but also guides the design of artificial cell polarization system with desired functions. The joint study on a simple 2-node network and *C. elegans* 5-node network demonstrated that the cell polarization pattern with both stability and asymmetry can be explicitly realized by a combinatorial network and spatially inhomogeneous parameters, which is expected to facilitate the interpretation of natural systems and design of artificial systems.

Although we deciphered the stability and asymmetry of the pattern generated by the *C. elegans* cell polarization network, additional experimental details are required to achieve a more comprehensive understanding of this system, as outlined below.

1) Although the model in this paper has tried to incorporate as many molecular species as possible for the maintenance phase of cell polarization, the actual network may include yet unidentified molecular species. This also extends to interactions, as well as their intensity, between molecular species. For instance, the components of the molecular complex (*e.g.*, PAR-3/PAR-6/PKC-3), often treated as a single entity in theoretical studies [Tostevin et al., *Biophys. J.*, 2008; Gross et al., *Nat. Phys.*, 2019; Seirin-Lee et al., *Cells*, 2020; Lim et al., *Cell Rep.*, 2021], might not be ideally identical concerning their slightly different polarized patterns [Wang et al., *Nat. Cell Biol.*, 2017]. The PAR-6 and PKC-3 also bind CDC-42, in addition to PAR-3, to mediate membrane association of CHIN-1 via phosphorylation, and such joint effect was previously described to be governed by a single node, CDC-42 [Lim et al., *Cell Rep.*, 2021]. Future experimental validation is needed to detail whether and how the molecular species described in Table S2 (and beyond) interact with each other.
2) The large number of molecular species introduces substantial parameter complexity, with myriad possible value assignments. Simplifying the model is necessary to manage computational costs unless all parameters can be measured accurately *in vivo* [Wang et al., *Nat. Commun.*, 2020].
3) Quantitative biological details, such as the concentration distributions, diffusions, productions, and degradations of molecular species on the cell membrane versus in the cell cytosol [Tostevin et al., *Biophys. J.*, 2008; Goehring et al., *Science*, 2011; Kravtsova et al., *Bull. Math. Biol.*, 2014; Sailer et al., *Dev. Cell*, 2015; Seirin-Lee, et al., *J. Theor. Biol.*, 2015; Seirin-Lee et al., *J. Theor. Biol.*, 2016; Gross et al., *Nat. Phys.*, 2019; Seirin-Lee et al., *J. Math. Biol.*, 2020; Lim et al., *Cell Rep.*, 2021] and the precise mathematical forms of regulatory interactions (*e.g.*, Hill equations or product equations) [Goehring et al., *Science*, 2011; Seirin-Lee et al., *Cells*, 2020; Lim et al., *Cell Rep.*, 2021], challenge model assumptions and hinder the unification of parameter value with unit between theory and experiment [Tostevin et al., *Biophys. J.*, 2008; Kravtsova et al., *Bull. Math. Biol.*, 2014]. Nonetheless, prior studies integrating theory and experiments have demonstrated that unveiling mathematical principles for regulatory networks does not strictly depend on unit unification; numerical network structure and parameter analysis can still effectively draw the mathematical principles for guiding the synthesis of artificial networks in reality, including the one for cell polarization [Elowitz et al., *Nature*, 2000; Gardner et al., *Nature*, 2000; Ma et al., *Cell*, 2009; Chau et al., *Cell*, 2012].

4) Incorporating biological processes across temporal stages and spatial scales could offer a more holistic understanding of cell polarization. On one hand, early stages, such as sperm entry and the establishment phase, which involve active actomyosin contractility and flow drive PAR-3/PAR-6/PKC-3 to be localized in the anterior, can provide insights into how initial polarized concentration distributions are achieved [Cuenca et al., *Development*, 2003; Motegi et al., *Nat. Cell Biol.*, 2011]. On the other hand, molecular (e.g., downstream localization of the myosin motor NMY-2 to the anterior cortex mediated by CDC-42 and PAR-2 [Goehring et al., *J. Cell Biol.*, 2011; Rose et al., *WormBook*, 2014; Geβele et al., *Nat. Commun.*, 2020]) and cellular (asymmetric segregation of cell volume and molecular content, mediated by molecular species like MES-1 and SRC-1 [Arata et al., *Development*, 2010; Seirin-Lee, *J. Theor. Biol.*, 2016]) effect of cell polarization can further elucidate its various roles in embryogenesis.

Beyond the *C. elegans* zygote studied in this paper, cell polarization also exists in later stages of *C. elegans* embryogenesis as well as in other organisms, where diverse functional dynamics have been reported [Rose et al., *WormBook*, 2014; Knoblich et al., *Nat. Rev. Mol. Cell Biol.*, 2001]. Governed by the cell polarization, the *C. elegans* embryo actually proceeds 4 rounds of asymmetric division to produce 4 somatic founder cells, then such reaction-diffusion pattern will lose its sharp transition plane as it is unscalable over cell size, leading to symmetric cell division at last [Hubatsch et al., *Nat. Phys.*, 2019]. Furthermore, as the reaction-diffusion network itself is not particularly constricted in the cellular scale, but also possibly in subcellular and multicellular scales, it would be interesting to investigate if the general principles proposed in this paper are also at work for the network programming and pattern modulation in other scales [Chao et al., *Cell*, 2014; Guan et al., *Sci. Adv.*, 2022].

Nowadays, the *de novo* construction of artificial cells with designated new functions is emerging, demanding theoretical guidance about how to synthesize the molecular control circuits beneath [Zhou et al., *Science*, 2023; Zhu et al., *Sci. Adv.*, 2023]. Like toggle switch, oscillation, and adaptation [Gardner et al., *Nature*, 2000; Zhou et al., *Science*, 2023; Sun et al., *Nucleic Acids Res.*, 2022], the polarized distribution of molecular species (accounting for cell polarization) is also another elementary behavior in a cell, and the corresponding circuits have been synthesized successfully around a decade ago [Chau et al., *Cell*, 2012]. However, it remains unclear whether the distribution pattern (*e.g.*, the transition plane of a molecular species) can be altered while maintaining stability. Taking the *C. elegans* network as an example, our computational framework has been demonstrated to be capable of deciphering such a real reaction-diffusion network with pattern stability and asymmetry, and has been packed as a user-friendly software, *PolarSim* (Supplemental Text); thus we believe it can be used for not only understanding the natural molecular networks reported before but also designing new molecular networks in the polarized cells with more physiological functions, for example, with tunable transition plane of molecular species to guide the allocation of downstream fate determinants and unequal cleavage during cytokinesis [Fickentscher et al., *Sci. Rep.*, 2017]. In the future, our computational framework for cell polarization network could be linked to the one about cytoskeleton activity, achieving a more comprehensive modeling description for cell division [Pavin et al., *New J. Phys.*, 2012; Ma et al., *New J. Phys.*, 2014].

## Conflict of interest

The authors declare that they have no conflict of interest.

## Acknowledgments

We thank Prof. Wei Wang, Dr. Daimin Li, Peng Xie, Shiyu Shen, and Qi Ding for their assistance in the numerical simulation, and Prof. Hongli Wang for his constructive advice on the project. We are grateful to Dr. Siyu Li, Dr. Baoshuang Shang, Dr. Shipu Xu, and Yuqing Zhong for their help during Yixuan Chen and Guoye Guan's academic visit to Songshan Lake Materials Laboratory. We appreciate Zhengyang Han for his assistance in improving the paper materials. Gratitude is extended to Prof. Zhongying Zhao for his help during Guoye Guan's academic visit to Hong Kong Baptist University. This work was supported by funding from the National Natural Science Foundation of China (12090053 and 32088101) and the Research Grants Council of the Hong Kong SAR (12303219). Computation was performed partly on the High-Performance Computing Platform at Peking University.

**Author contributions**

Y.C. and G.G. designed and conceived the project, and wrote the paper; Y.C., G.G., L.-H.T., and C.T. performed research and conducted analysis; L.-H.T. and C.T. supervised the project and revised the paper.

**References**

[1] Aceto D, Beers M, Kemphues KJ. Interaction of PAR-6 with CDC-42 is required for maintenance but not establishment of PAR asymmetry in *C. elegans*. Dev Biol 2006;299:386-97.

[2] Ajduk A, Zernicka-Goetz M. Polarity and cell division orientation in the cleavage embryo: from worm to human. Mol Hum Reprod 2016;22:691-703.

[3] Arata Y, Lee JY, Boldstein B, et al. Extracellular control of PAR protein localization during asymmetric cell division in the *C. elegans* embryo. Development 2010;137:3337-45.

[4] Beatty A, Morton D, Kemphues K. The *C. elegans* homolog of *Drosophila* lethal giant larvae functions redundantly with PAR-2 to maintain polarity in the early embryo. Development 2010;137:3995-4004.

[5] Beatty A, Morton DG, Kemphues K. PAR-2, LGL-1 and the CDC-42 GAP CHIN-1 act in distinct pathways to maintain polarity in the *C. elegans* embryo. Development 2013;140:2005-14.

[6] Bilder D, Li M, Perrimon N. Cooperative regulation of cell polarity and growth by *Drosophila* tumor suppressors. Science 2000;289:113-6.

[7] Blanchoud S, Busso C, Naef F, et al. Quantitative analysis and modeling probe polarity establishment in *C. elegans* embryos. Biophys J 2015;108:799-809.

[8] Chao JT, Wong AKO, Tavassoli S, et al. Polarization of the endoplasmic reticulum by ER-septin tethering. Cell 2014;158:620-32.

[9] Chau AH, Walter JM, Gerardin J, et al. Designing synthetic regulatory networks capable of self-organizing cell polarization. Cell 2012;151:320-32.

[10] Cuenca AA, Schetter A, Aceto D, et al. Polarization of the *C. elegans* zygote proceeds via distinct establishment and maintenance phases. Development 2003;130:1255–65.

[11] Doe CQ, Bowerman B. Asymmetric cell division: fly neuroblast meets worm zygote. Curr Opin Cell Biol 2001;13:68-75.

[12] Elowitz MB, Leibler S. A synthetic oscillatory network of transcriptional regulators. Nature 2000;403:335-8.

[13] Etienne-Manneville S, Hall A. Rho GTPases in cell biology. Nature 2002;420:629-35.

[14] Fickentscher R, Krauss SW, Weiss M. Anti-correlation of cell volumes and cell-cycle times during the embryogenesis of a simple model organism. New J Phys 2018;20:113001.

[15] Fickentscher R, Weiss M. Physical determinants of asymmetric cell divisions in the early development of *Caenorhabditis elegans*. Sci Rep 2017;7:9369.

[16] Fickentscher R, Struntz P, Weiss M. Setting the clock for fail-safe early embryogenesis. Phys Rev Lett 2016;117:188101.

[17] Fishman GS. Monte Carlo: Concepts, algorithms, and applications. New York, NY: Springer. 1995.

[18] Gardner TS, Cantor CR, Collins JJ. Construction of a genetic toggle switch in *Escherichia coli*, Nature 2000;403:339-42.

[19] Geßele R, Halatek J, Würthner L, et al. Geometric cues stabilise long-axis polarisation of PAR protein patterns in *C. elegans.* Nat Commun 2020;11:539.

[20] Goehring NW, Hoege C, Grill SW, et al. PAR proteins diffuse freely across the anterior-posterior boundary in polarized *C. elegans* embryos. J Cell Biol 2011;193:583-94.

[21] Goehring NW, Trong PK, Bois JS, et al. Polarization of PAR proteins by advective triggering of a pattern-forming system. Science 2011;334:1137-41.

[22] Goldstein B, Hird SN. Specification of the anteroposterior axis in *Caenorhabditis elegans*. Development 1996;122:1467-74.

[23] Goodrich LV, Strutt D. Principles of planar polarity in animal development. Development 2011;138:1877-92.

[24] Gotta M, Abraham MC, Ahringer J. CDC-42 controls early cell polarity and spindle orientation in *C. elegans*. Curr Biol 2001;11:482-8.

[25] Gross P, Kumar KV, Goehring NW, et al. Guiding self-organized pattern formation in cell polarity establishment. Nat Phys 2019;15:293-300.

[26] Guan C, Qiao L, Xiong Y, et al. Coactivation of antagonistic genes stabilizes polarity patterning during shoot organogenesis. Sci Adv 2022;8:eabn0368.

[27] Guan G, Wong MK, Zhao Z, et al. Volume segregation programming in a nematode's early embryogenesis. Phys Rev E 2021;104:054409.

[28] Hoege C, Constantinescu A, Schwager A, et al. LGL can partition the cortex of one-cell *Caenorhabditis elegans* embryos into two domains. Curr Biol 2010;20:1296-303.

[29] Hubatsch L, Peglion F, Reich JD, et al. A cell-size threshold limits cell polarity and asymmetric division potential. Nat Phys 2019;15:1075-85.

[30] Jankele R, Jelier R, Gönczy P. Physically asymmetric division of the *C. elegans* zygote ensures invariably successful embryogenesis. eLife 2021;10:e61714.

[31] Kemphues KJ, Priess JR, Morton DG, et al. Identification of genes required for cytoplasmic localization in early *C. elegans* embryos. Cell 1988;52:311-20.

[32] Kim M, Datta A, Brakeman P, et al. Polarity proteins PAR6 and aPKC regulate cell death through GSK-3β in 3D epithelial morphogenesis. J Cell Sci 2007;120:2309-17.

[33] Knoblich JA. Asymmetric cell division during animal development. Nat Rev Mol Cell Biol 2001;2:11-20.

[34] Kondo Y, Ledderose C, Slubowski CJ, et al. Frontline Science: *Escherichia coli* use LPS as decoy to impair neutrophil chemotaxis and defeat antimicrobial host defense. J Leukoc Biol 2019;106:1211-9.

[35] Koorman T, Klompstra D, van der Voet M, et al. A combined binary interaction and phenotypic map of *C. elegans* cell polarity proteins. Nat Cell Biol 2016;18:337-46.

[36] Kravtsova N, Dawes AT. Actomyosin regulation and symmetry breaking in a model of polarization in the early *Caenorhabditis elegans* embryo. Bull Math Biol 2014;76:2426-48 (2014).

[37] Kumfer KT, Cook SJ, Squirrell JM, et al. CGEF-1 and CHIN-1 regulate CDC-42 activity during asymmetric division in the *Caenorhabditis elegans* embryo. Mol Biol Cell 2010;21:266-77.

[38] Lang CF, Munro E. The PAR proteins: from molecular circuits to dynamic self-stabilizing cell polarity. Development 2017;144:3405-16.

[39] Lin DW, Liu Y, Lee YQ, et al. Construction of intracellular asymmetry and asymmetric division in *Escherichia coli*. Nat Commun 2021;12:888.

[40] Lim YW, We FL, Shankar P, et al. A balance between antagonizing PAR proteins specifies the pattern of asymmetric and symmetric divisions in *C. elegans* embryogenesis. Cell Rep 2021;36:109326.

[41] Ma R, Laan L, Dogterom M, et al. General theory for the mechanics of confined microtubule asters. New J Phys 2014;16:013018.

[42] Ma W, Trusina A, El-Samad H, et al. Defining network topologies that can achieve biochemical adaptation. Cell 2009;138:760-73.

[43] Macara IG. Parsing the polarity code. Nat Rev Mol Cell Biol 2004;5:220-31.

[44] The MathWorks Inc. MATLAB version: 9.13.0 (R2022b), Natick, Massachusetts: The

MathWorks Inc. 2022.

[45] Morita Y, Seirin-Lee S. Long time behavior and stable patterns in high-dimensional polarity models of asymmetric cell division. J Math Biol 2021:82:66.

[46] Motegi F, Zonies S, Hao Y, et al. Microtubules induce self-organization of polarized PAR domains in *Caenorhabditis elegans* zygotes. Nat Cell Biol 2011;13:1361–7.

[47] Munro E, Nance J. Cortical flows powered by asymmetrical contraction transport PAR proteins to establish and maintain anterior-posterior polarity in the early *C. elegans* embryo. Dev Cell 2004;7:413-24.

[48] Nance J, Getting to know your neighbor: cell polarization in early embryos. J Cell Biol 2014;206:823-32.

[49] Pavin N, Laan L, Ma R, et al. Positioning of microtubule organizing centers by cortical pushing and pulling forces. New J Phys 2012; 14:105025.

[50] Rose LS, Gönczy P. Polarity establishment, asymmetric division and segregation of fate determinants in early *C. elegans* embryo. WormBook 2014;1-43.

[51] Sailer A, Anneken A, Li Y, et al. Dynamic opposition of clustered proteins stabilizes cortical polarity in the *C. elegans* zygote. Dev Cell 2015;35:131-42.

[52] Schneider SQ, Bowerman B. Cell polarity and the cytoskeleton in the *Caenorhabditis elegans* zygote. Annu Rev Genet 2003;37:221-49.

[53] Schubert CM, Lin R, de Vries CJ, et al. MEX-5 and MEX-6 function to establish soma/germline asymmetry in early *C. elegans* embryos. Mol. Cell 2000:5:671-82.

[54] Seirin-Lee S. Asymmetric cell division form a cell to cells: shape, length, and location of polarity domain. Dev Growth Differ 2020;62:188-95.

[55] Seirin-Lee S. Positioning of polarity formation by extracellular signaling during asymmetric cell division. J Theor Biol 2016;400:52-64.

[56] Seirin-Lee S, Shibata T. Self-organization and advective transport in the cell polarity formation for asymmetric cell division. J Theor Biol 2015;382:1-14.

[57] Seirin-Lee S, Gaffney EA, Dawes AT. CDC-42 interactions with Par proteins are critical for proper patterning in polarization. Cells 2020;9:2036.

[58] Seirin-Lee S. The role of cytoplasmic MEX-5/6 polarity in asymmetric cell division. Bull Math Biol 2021;83:29.

[59] Seirin-Lee S, Sukekawa T, Nakahara T, et al. Transitions to slow or fast diffusions provide a general property for in-phase or anti-phase polarity in a cell. J Math Biol 2020; 80:1885-917.

[60] Sulston JE, Schierenberg E, White JG, et al. The embryonic cell lineage of the nematode

*Caenorhabditis elegans*. Dev Biol 1983;100:64-119.

[61] Sun Z, Wei W, Zhang M, et al. Synthetic robust perfect adaptation achieved by negative feedback coupling with linear weak positive feedback. Nucleic Acids Res 2022;50:2377-86.

[62] Tian B, Guan G, Tang L-H, et al. Why and how the nematode's early embryogenesis can be precise and robust: a mechanical perspective. Phys Biol 2020;17:026001.

[63] Tostevin F, Howard M. Modeling the establishment of PAR protein polarity in the one-cell *C. elegans* embryo. Biophys J 2008;95:4512-22.

[64] Wang JT, Seydoux G. Germ cell specification. Adv Exp Med Biol 2013;757:17-39.

[65] Wang S-C, Low TYF, Nishimura Y, et al. Cortical forces and CDC-42 control clustering of PAR proteins for *Caenorhabditis elegans* embryonic polarization. Nat Cell Biol 2017;19:988-95.

[66] Wang T, You L. The persistence potential of transferable plasmids. Nat Commun 2020;11:5589.

[67] Watson JL, Krüger LK, Ben-Sasson AJ, et al. Synthetic Par polarity induces cytoskeleton asymmetry in unpolarized mammalian cells. Cell 2023;186:4710-27.

[68] Zhou Z, Liu Y, Feng Y, et al. Engineering longevity—design of a synthetic gene oscillator to slow cellular aging. Science 2023;380:3760381.

[69] Zhu R, Santat LA, Markson JS, et al. Reconstitution of morphogen shuttling circuits. Sci Adv 2023;9:eadf9336.